\documentclass[10pt,twocolumn,twoside]{IEEEtran}
\usepackage{latexsym}
\usepackage{amsfonts}
\usepackage{amsbsy}
\usepackage{amsmath,amssymb}
\usepackage{times}
\usepackage{graphicx}
\usepackage{stkernel}
\usepackage{enumerate}
\usepackage[usenames]{color}
\usepackage[dvips]{pstcol}
\usepackage{epstopdf}
\input epsf
\title{Robust MMSE Precoding for Multiuser MIMO Relay Systems using Switched Relaying and Side Information}

 \author{Yunlong Cai, Rodrigo C. de Lamare, Lie-Liang Yang, and Minjian Zhao
\thanks{

Copyright (c) 2013 IEEE. Personal use of this material is permitted.
However, permission to use this material for any other purposes must
be obtained from the IEEE by sending a request to
pubs-permissions@ieee.org.

Part of this work was presented at WCNC2013 \cite{yunlongwcnc2013}.

Y. Cai and M. Zhao are with Department of Information Science and
Electronic Engineering, Zhejiang University, Hangzhou 310027, China
(e-mail: ylcai@zju.edu.cn; mjzhao@zju.edu.cn).

R. C. de Lamare is with CETUC-PUC-Rio, 22453-900 Rio de Janeiro,
Brazil, and also with the Communications Research Group, Department
of Electronics, University of York, York Y010 5DD, U.K. (e-mail:
rcdl500@ohm.york.ac.uk).

L.-L. Yang is with the School of Electronics and Computer Science,
University of Southampton, SO17 1BJ Southampton, U.K.
(e-mail:lly@ecs.soton.ac.uk).

This work was supported in part by the National Science Foundation
of China (NSFC) under Grant $61101103$, the Fundamental Research
Funds for the Central Universities and the Scientific Research
Project of Zhejiang Provincial Education Department.

 } }

\begin{document}
\maketitle \thispagestyle{empty}

\vspace{-2.3em}

\begin{abstract}
This study proposes a novel precoding scheme for multiuser multiple-input multiple-output (MIMO) relay systems in the presence of imperfect channel state information (CSI).
 The base station (BS) and the MIMO relay station (RS) are both equipped with the same codebook of unitary matrices.
 According to each element of the codebook, we create a \textit{latent} precoding matrix pair, namely a BS precoding matrix and an RS precoding matrix.
The RS precoding matrix is formed by multiplying the appropriate unitary matrix from the codebook by a power scaling factor.
 Based on the given CSI and a block of transmit symbols, the optimum precoding matrix pair, within the class of all possible latent precoding matrix pairs derived from the various unitary matrices,
  is selected by a suitable  selection  mechanism for transmission, which is designed to minimize the squared Euclidean distance between the pre-estimated  received vector and the true transmit symbol vector. We develop a minimum mean square error (MMSE) design algorithm for the construction of the latent precoding matrix pairs.
   In the proposed scheme, rather than sending the complete  processing matrix, only the index of the unitary matrix and its power scaling factor are sent by the BS to the RS. This significantly reduces the overhead.
Simulation results show that compared to other recently reported precoding algorithms the proposed  precoding scheme is capable of providing improved robustness against the effects of CSI estimation errors and multiuser interference.
\\


\end{abstract}

\IEEEpeerreviewmaketitle

\section{Introduction}

%
 Optimum non-regenerative relay station (RS)  precoding matrices for single user MIMO relay systems have been well studied in the literature \cite{WGUAN}-\cite{fanshuotseng2}.
%
 Guan and Luo  employed the constrained minimum mean squared error
(MMSE) criterion to jointly design the RS precoding matrix
and the  receive filtering matrix at the destination \cite{WGUAN}.
In \cite{yuerong1}, Rong and Gao generalized the optimum RS precoding matrix by including the direct link.
In the case of imperfect channel state information (CSI),
 Xing \emph{et
al.} \cite{chengwenxing1} proposed a joint robust design algorithm for the linear  RS precoding matrix and
the  receive filtering matrix.
By taking base station (BS) precoding into account   some researchers investigated the
joint design algorithm of the BS and RS precoding matrices
\cite{Binzhang2}-\cite{fanshuotseng2}.
 In particular, Zhang \emph{et
al.}  \cite{Binzhang2} proposed a
joint iterative optimization algorithm to design the BS
precoding matrix,  RS precoding matrix and    receive filtering matrix.
 Tseng and Wu designed an iterative algorithm by
minimizing the MMSE upper bound, instead of the original MMSE in \cite{fanshuotseng}, \cite{fanshuotseng2}.

 Of recent, the study of precoding techniques in multiuser MIMO relay systems is becoming more and more of importance
 \cite{Rzhang}-\cite{zijianwang}.
  Zhang \emph{et
al.} \cite{Rzhang} minimized the weighted sum-power consumption under the minimum
quality-of-service (QoS) constraints by  jointly optimizing
linear beamforming and power control at the BS and RS.
  Chae \emph{et
al.} \cite{Njindal}
combined the BS nonlinear precoding matrix with a linear processing algorithm
at the RS.  They also relied on the fact  that the CSI can be obtained via feedback or channel reciprocity.
  By using feedback quantized CSI, while  assuming  perfect CSI at the destination, two robust linear precoding schemes at the RS based on zero forcing (ZF)  and the MMSE criteria were proposed for downlink multiuser MIMO relay systems  to handle  quantization errors, \cite{Binzhang}.  However, the authors did not  consider  BS precoding  in their work.
 In \cite{weixu1}, Xu \emph{et al.} proposed singular value decomposition (SVD) based  BS and RS precoding algorithms under the assumption of perfectly known CSI, where the BS precoding matrix is designed based on the SVD of the first time slot channel, and the ZF RS precoding matrix is obtained based on the second time slot channel.
 In the presence of imperfect CSI, the studies in \cite{weixu2} and \cite{zijianwang} proposed  extensions of the method reported in \cite{weixu1},  where  robust precoding algorithms were developed to deal with  CSI quantization and estimation errors, respectively.
 Notice that the BS and RS precoding matrices in \cite{ weixu1}-\cite{zijianwang} are not designed to optimize a cost function regarding an overall system performance, therefore  they may suffer from performance degradation.

In this work, we propose a novel precoding scheme based on switched relaying (SR) processing
for multiuser MIMO relay systems.
In practice,  in cellular systems it is preferable  to implement most of the signal processing operations at the BS rather than at the RS, since the BS is more powerful and the RS is expected to have a simple structure and low energy consumption \cite{Dfeng1}-\cite{mBanerjee}.
In this regard, the proposed technique
is implemented at the BS.  The BS and the MIMO RS are both equipped with the same codebook of unitary matrices.
 According to each element of the codebook, we create a \textit{latent} precoding matrix pair, namely a BS precoding matrix and an RS precoding matrix.
 The RS precoding matrix is formed by multiplying the appropriate unitary matrix from the codebook by a power scaling factor. We develop a design algorithm for computing the BS precoding matrix and RS power scaling factor  in order to construct the latent precoding matrix pairs.
Based on the given CSI and  a block of transmitted symbols,
the optimum pair within the  group of latent precoding matrix pairs is chosen by a suitable  selection  mechanism for transmission, which is designed to minimize the squared Euclidean distance between the noiseless pre-estimated  received vector and the true transmit symbol vector.
 Prior to  payload transmission, the BS  transmits the index of the unitary matrix and   the RS power scaling factor information corresponding to the optimum latent precoding matrix pair to the RS through signalling channels \cite{vook}-\cite{CBchae}, where the optimum RS precoding matrix is formed\footnote{Rather than sending the  complete processing matrix, the proposed scheme only sends forward limited information to the RS  from the BS  which significantly reduces the overhead.}. In addition, we propose a method based on the most frequently selected candidates (MSC) for the codebook design. An analysis of the proposed algorithm in terms of computational complexity, probability of error and requirement of side information is carried out. Simulation results demonstrate that  the proposed SR-based precoding scheme is capable of providing improved robustness against the effects of CSI estimation errors and interference compared to the existing precoding algorithms.

 This paper is organized as follows. Section \ref{Section2:system}
briefly describes the system model.
The proposed SR-based transmission scheme is introduced
in
Section \ref{Section3:perfectcsi} in terms of the latent precoding matrix design algorithm,
 the  selection  mechanism of the optimization latent precoding matrix pair and the codebook design.
 An  analysis  of the proposed algorithm  is conducted in Section \ref{Section5:analysis2}.
 Simulation results are presented
in Section \ref{Section6:simulations} and finally conclusions are drawn in  Section
\ref{Section7:conclusion}.

In this paper, the superscripts $(.)^{T}$, $(.)^{*}$, $(.)^{-1}$, and $(.)^{H}$
denote transpose, element-wise conjugate, matrix inverse,
 and Hermitian transpose, respectively.
Bold symbols denote matrices or vectors. The symbols $E[.]$, $|.|$,
$||.||$,  ${\it Tr}\{.\}$
and $\mathbf{I}$ represent the  expectation operator, the norm
of a scalar, the norm of a vector,  the
trace operation of a square matrix and an identity
matrix of appropriate dimension, respectively.
The operation $(x,:)$ denotes taking the $x$-th row vector
from a matrix. The operation $(:, y)$ denotes taking the $y$-th column vector from a matrix.
$\Re[.]$ selects the real part. $[y]^{+}=\max [0,y]$. $||.||_{F}$ denotes the matrix Frobenius norm. The factor $\otimes$ denotes the operation of the Kronecker
product.

\section{System model}
\label{Section2:system}

We consider the downlink of a multiuser MIMO cellular system consisting
of one BS, one  RS, and $K$ mobile stations (MSs).
In practice, this model is employed for the relay architectures of
3GPP LTE-Advanced \cite{swpeters}.
 We consider a case in which
 the BS and RS
are equipped with $N_{t}$ and $N_{r}$ antennas, respectively, and the MS is equipped with a
single antenna, where $K\leq \min\{N_{t},N_{r}\}$.
 In addition,  the BS
and the RS are equipped with a finite codebook of  unitary matrices,
 i.e. $\mathcal{T}=\{\mathbf{T}_1,\mathbf{T}_2,\ldots,\mathbf{T}_{2^B}\}$, $2^{B}$ is the codebook size, and
 consider
half time-division duplex (TDD) non-regenerative relaying \cite{Laneman2}.
 In the first phase,
the received vector at the RS is given by
\begin{equation}
\mathbf{r}^{(l)}_{R}=\mathbf{H}_{1}\mathbf{P}_{l}\mathbf{b}+\mathbf{n}_{1}.
\end{equation}
 In this expression,
 the elements of transmit symbol vector $\mathbf{b}=[b_{1},  \ldots, b_{K}]^{T}\in \mathcal{C}^{K\times 1}$ are independent and identically distributed (i.i.d.), $E[|b_{k}|^2]=1$, $k \in \{1,\ldots,K\}$, and
$\mathbf{P}_{l}\in \mathcal{C}^{N_{t}\times K}$
denotes the   BS precoding matrix corresponding to the $l$-th latent precoding matrix pair, where
$\forall l \in \{1,\ldots,2^{B}\}$.
The matrix $\mathbf{H}_{1}\in \mathcal{C}^{N_{r}\times N_{t}}$ is the  channel matrix between the BS and the RS,
 whose
elements are  i.i.d.
complex circular Gaussian variables with zero mean and unit variance, which we indicate by the standard notation $\mathcal{CN}\left(0,1 \right)$,
and
$\mathbf{n}_{1}\in \mathcal{C}^{N_{r}\times 1}$ is the  additive complex Gaussian
noise with covariance matrix
$E[\mathbf{n}_{1}\mathbf{n}^{H}_{1}]=\mathbf{\sigma}^{2}_{1}\mathbf{I}$, where $\sigma^{2}_{1}$ denotes the first phase noise variance.

 In the second phase, the vector $\mathbf{r}^{(l)}_{R}\in \mathcal{C}^{N_{r}\times N_{r}}$ is operated
by the   RS precoding matrix
$\mathbf{W}_{l}$ corresponding to the $l$-th latent precoding matrix pair, where $\mathbf{W}_l$ is formed by multiplying the appropriate unitary matrix from the codebook by a power scaling factor, i.e.
$\mathbf{W}_{l}=\beta_{l}\mathbf{T}_{l}$, and  $\mathbf{T}_{l}$ and $\beta_{l}$ denote the selected unitary matrix from the codebook and the RS power scaling factor, respectively.
The forwarded signal vector from the RS is given
by
\begin{equation}
\mathbf{x}^{(l)}_{R}=\mathbf{W}_{l} \mathbf{r}^{(l)}_{R}.
\end{equation}
The BS and RS power constraints are $E[||\mathbf{P}_{l}\mathbf{b}||^{2}]= {\it Tr}\{\mathbf{P}_{l}\mathbf{P}^{H}_{l}\}\leq P_{t}$ and $E[||\mathbf{x}^{(l)}_{R}||^{2}]=\beta^{2}_{l}E\Big[ {\it Tr}\{\mathbf{T}_{l}(\mathbf{H}_{1}\mathbf{P}_{l}\mathbf{P}^{H}_{l}\mathbf{H}^{H}_{1}+\sigma^{2}_{1}\mathbf{I})\mathbf{T}^{H}_{l}\}\Big]=\beta^{2}_{l} E\Big[ {\it Tr}\{\mathbf{H}_{1}\mathbf{P}_{l}\mathbf{P}^{H}_{l}\mathbf{H}^{H}_{1}+\sigma^{2}_{1}\mathbf{I}\}\Big]\leq P_{r}$, respectively.
For the second phase, we model the transmission from the RS to the
$K$ MSs as a MIMO broadcast channel, we stack all the MSs'
received data and obtain the received vector
\begin{equation}
\begin{split}
\mathbf{y}^{(l)}&=\mathbf{H}_{2}\mathbf{x}^{(l)}_{R}+\mathbf{n}_{2}
=\mathbf{H}_{2}\mathbf{W}_{l}\mathbf{H}_{1}\mathbf{P}_{l}\mathbf{b}+\mathbf{H}_{2}\mathbf{W}_{l}\mathbf{n}_{1}+\mathbf{n}_{2},
\end{split}\label{eq:srsystemmodel}
\end{equation}
where $\mathbf{H}_{2}\in \mathcal{C}^{K\times N_{r}}$ is the  channel matrix
between the RS and the MSs, its entries are i.i.d. zero mean complex circular
Gaussian variables with unit variance, and $\mathbf{n}_{2}$ denotes
the additive complex Gaussian noise,
$E[\mathbf{n}_{2}\mathbf{n}^{H}_{2}]=\sigma^{2}_{2}\mathbf{I}$, where $\sigma^{2}_{2}$ denotes the second phase noise variance.
 In order to obtain the required CSI,  we  simplify the two-hop channel estimation problem to two independent MIMO channel estimation procedures.
 The CSI can be estimated  at the BS and  RS, respectively, by using a specific channel estimation algorithm \cite{musavian}-\cite{Tseng2013tvt}. Since we consider the TDD mode in this work, the downlink transmit CSI can be obtained  due to  reciprocity \cite{vook}.
  Note that the RS needs to feed  the estimated second phase CSI back to the BS by using signalling channels \cite{vook}-\cite{CBchae}. In this work, we assume that  the channel varies sufficiently slowly, and  the BS can obtain the estimated second phase CSI.
To model the
statistical distribution of the estimation errors in the channel matrices,
the well-known Kronecker model is
adopted here for the covariance matrix of the CSI mismatch \cite{chengwenxing1}.
In particular, the true (but unknown) channel matrix is expressed as follows,
\begin{equation}
\mathbf{H}_{j}=\mathbf{\hat{H}}_{j}+\Delta\mathbf{H}_{j}, \quad j=1,2, \label{eq:CSIerror1}
\end{equation}
where $\mathbf{\hat{H}}_{j}$ denotes the
estimated channel matrices, while $\Delta\mathbf{H}_{j}$ denotes the corresponding channel estimation
error matrix. The latter can be expressed as
\begin{equation}
\Delta\mathbf{H}_{j}=\mbox{\boldmath$\Sigma$}^{\frac{1}{2}}_{j}\mathbf{\tilde{H}}_{j}\mbox{\boldmath$\Psi$}^{\frac{1}{2}}_{j},
\end{equation}
where the elements of $\mathbf{\tilde{H}}_{j}$ are i.i.d. Gaussian
random variables with zero mean and unit variance, and
$\mbox{\boldmath$\Psi$}_{j}$ and $\mbox{\boldmath$\Sigma$}_{j}$
denote the covariance matrices of the  channel seen from
the transmitter and receiver, respectively. Furthermore, the matrix
$\Delta\mathbf{H}_{j}$ has the matrix-variate complex circular Gaussian
distribution, which can be expressed as $\Delta\mathbf{H}_{j}\sim \mathcal{CN}\left(\mathbf{0}, \mbox{\boldmath$\Sigma$}_{j} \otimes \mbox{\boldmath$\Psi$}^{T}_{j}\right)$ \cite{chengwenxing1},
\cite{james}, \cite{cxing2}.
By using the estimation algorithm in \cite{musavian}, we have $\mbox{\boldmath$\Psi$}_{j}=\mathbf{R}_{T,j}$ and $\mbox{\boldmath$\Sigma$}_{j}=\sigma^{2}_{e,j}\mathbf{R}_{R,j}$, where
 $\mathbf{R}_{T,j}$ and $\mathbf{R}_{R,j}$ are the  transmit and receive antennas correlation matrices, respectively, and $\sigma^{2}_{e,j}$ is the   channel estimation error variance.
It is reasonable to assume that $\mbox{\boldmath$\Psi$}_{j}$ and $\mbox{\boldmath$\Sigma$}_{j}$  are slowly
varying and can be known a priori by estimating long term channel statistics. It is important to note that the analysis to be presented in this paper can be applied in exactly the same way
without assuming any specific form  for the matrices $\mbox{\boldmath$\Psi$}_{j}$ and $\mbox{\boldmath$\Sigma$}_{j}$ as long as they are symmetric and full-rank \cite{musavian}, \cite{ding2}.

\section{Proposed SR-based Precoding Scheme}
\label{Section3:perfectcsi}

\begin{figure*}[!t]
\centering \scalebox{0.54}{\includegraphics{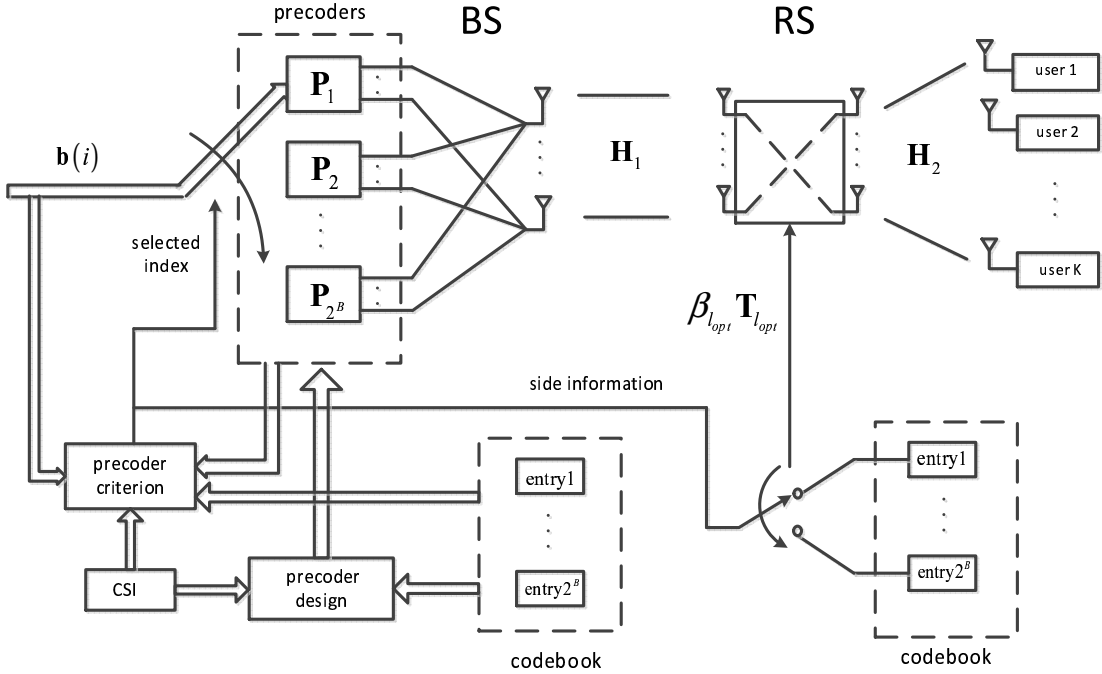}}
\caption{ Proposed  precoding scheme using switched relaying processing for multiuser MIMO relay systems. } \label{fig:mimorelay}
\end{figure*}

As shown in
Fig. \ref{fig:mimorelay},  each unitary matrix in the codebook gives rise to a latent precoding matrix pair, namely a  BS precoding matrix and an RS precoding matrix. The RS precoding matrix is formed by multiplying the appropriate codebook unitary matrix by a power scaling factor.
Considering the size $2^B$ of the codebook, we can therefore design $2^B$ such latent  precoding matrix pairs  corresponding to each unitary matrix. The proposed SR-based relay transmission scheme works as follows.
\begin{enumerate}[$\bullet$]
\item  For  the given first and second phase channel conditions (CSI), i.e. $\mathbf{\hat{H}}_1$ and $\mathbf{\hat{H}}_2$:
\begin{enumerate}[$\bullet$]
\item The BS  computes all the $2^B$ latent precoding matrix pairs (BS and RS precoding matrices)
based on all the entries in the codebook of  unitary matrices\footnote{We use unitary matrices as the codebook entries  because each unitary matrix can generate an equivalent channel matrix $\bar{\mathbf{H}}_{l}=\mathbf{H}_{2}\mathbf{T}_{l}\mathbf{H}_{1}$ by performing  rotations, and $2^B$ latent precoding matrix pairs produced by these equivalent channel matrices  provide different capabilities of interference cancellation and symbol detection. The optimum latent precoding matrix pair can be selected from them.}
and the estimated CSI.
\item For each transmission data block:
 \begin{enumerate}[1)~]
\item  Prior to transmission, the BS precodes the transmit data block with the BS precoding matrix taken from the computed optimum latent precoding matrix pairs.

\item  The BS transmits the index of the   unitary matrix and the   RS power scaling factor information corresponding to the optimum latent precoding matrix pair to the RS through a signalling channel.

\item  The RS determines the appropriate power scaling factor and then selects a unitary matrix   from the codebook based on the feedforward side information, and form the optimum RS precoding matrix.

\item The block of transmit symbols is transmitted based on the   BS precoding matrix and  retransmitted by the RS using the RS precoding matrix corresponding to the optimum latent precoding matrix pair.
\end{enumerate}
\end{enumerate}
\end{enumerate}

In this section,  we firstly describe the design algorithm for  the construction of the latent precoding matrix pairs.
 Secondly,
in order to choose the optimum latent precoding matrix pair before transmission, we propose a selection mechanism based on
the squared Euclidean distance.
 Finally,  the design method
for the codebook of  unitary matrices is described.

\subsection{Design for the Latent Precoding Matrix Pair}

Each latent precoding matrix pair is designed based on the equivalent channel matrix $\mathbf{\bar{H}}_{l}=\mathbf{H}_{2}\mathbf{T}_{l}\mathbf{H}_{1}$ corresponding to the $l$-th unitary matrix within the codebook $\mathcal{T}$.
 In order to construct the $2^B$ latent precoding matrix pairs, we need to compute the BS precoding matrix $\mathbf{P}_{l}$ and the RS power scaling factor $\beta_{l}$ for each latent precoding matrix pair.
  Remark that the RS precoding matrix $\mathbf{W}_{l}$ corresponding to the $l$-th latent precoding matrix pair consists of the $l$-th unitary matrix $\mathbf{T}_{l}$  and the RS power scaling factor $\beta_{l}$, namely $\mathbf{W}_l=\beta_l\mathbf{T}_l$.
 In the following, we propose a   design method by minimizing the MSE in the scenario of imperfect CSI.
 We aim to design the precoding matrix $\mathbf{P}_{l}$ and the RS power scaling factor $\beta_{l}$, which minimize the total MSE under the BS and RS power transmit power constraints.
The optimization problem is given by
\begin{equation}
\min_{\mathbf{P}_{l}, \beta_{l}}  \zeta
\end{equation}
\begin{equation}
{\rm s.t.} \quad  {\it Tr}\{\mathbf{P}_{l}\mathbf{P}^{H}_{l}\}= P_{t},
 \quad \beta^{2}_{l} E\Big[
{\it Tr}\{\mathbf{H}_{1}\mathbf{P}_{l}\mathbf{P}^{H}_{l}\mathbf{H}^{H}_{1}+\sigma^{2}_{1}\mathbf{I}\}\Big]\leq
P_{r},
\end{equation}
where\footnote{ In this work, a pre-fixed receiver is used at the destination to reduce complexity. In this case, the design of receiver only depends on the channel and is oblivious to transmitter \cite{Palomar2010}. Since we focus on the multiuser scenario with a single antenna, the pre-fixed receiver of each user should be a scalar and the virtual multi-antenna receive filtering matrix is an identity matrix \cite{Palomar2010}.  As a result, the attenuation and the phase shift from the BS to the mobile station can be pre-equalized by the proposed BS and RS precoding matrices. }
\begin{equation}
\begin{split}
\zeta&=E[||\mathbf{b}-\mathbf{y}^{(l)}||^2]\\&={\it Tr}\bigg\{E\bigg[\mathbf{I}-\beta_{l}\mathbf{P}^{H}_{l}\mathbf{\bar{H}}^{H}_{l}-\beta_{l}\mathbf{\bar{H}}_{l}\mathbf{P}_{l}+\beta^{2}_{l}\mathbf{\bar{H}}_{l}\mathbf{P}_{l}\mathbf{P}^{H}_{l}\mathbf{\bar{H}}^{H}_{l}\\&\quad+\frac{ {\it Tr}\{\mathbf{P}_{l}\mathbf{P}^{H}_{l}\}}{P_{t}}\beta^{2}_{l}(\sigma^{2}_{1}\mathbf{\hat{H}}_{2}\mathbf{\hat{H}}^{H}_{2}+\sigma^{2}_{1}\Delta\mathbf{H}_{2}\Delta\mathbf{H}^{H}_{2})
\\&\quad+\sigma^{2}_{2}\frac{ {\it Tr}\{\mathbf{P}_{l}\mathbf{P}^{H}_{l}\}}{P_{t}}\mathbf{I}\bigg]\bigg\}.
 \end{split}
 \end{equation}
Note that $\mathbf{b}$, $\mathbf{n}_{1}$, $\mathbf{n}_{2}$, $\Delta\mathbf{H}_{1}$ and $\Delta\mathbf{H}_{2}$ are uncorrelated and we take expectation over them individually.
We used the fact that $\mathbf{T}_{l}$ is a unitary matrix and assume that $ {\it Tr}\{\mathbf{P}_{l}\mathbf{P}^{H}_{l}\}=P_{t}$ without loss of generality \cite{ADDabbagh}.
By employing the statistical property of the CSI estimation error (\ref{eq:CSIerror1}) and   applying the Karush-Kuhn-Tucker (KKT) conditions \cite{Sboyd},  we obtain the expressions for the robust BS  precoding matrix and the RS power scaling factor corresponding to the $l$-th latent precoding matrix pair as follows:
\begin{equation}
\begin{split}
\mathbf{P}_{l}&=\beta_{l}\Bigg(\mathbf{M}+\bigg(\frac{
\beta^{2}_{l}\sigma^{2}_{1} {\it Tr}\{\mathbf{\hat{H}}_{2}\mathbf{\hat{H}}^{H}_{2}\}}{P_{t}}+\frac{K\sigma^{2}_{2}}{P_{t}} \bigg)\mathbf{I}\\&\quad+\frac{
\beta^{2}_{l}
\sigma^{2}_{1} {\it Tr}\{\mbox{\boldmath$\Psi$}_{2}\}\mbox{\boldmath$\Sigma$}_{2}}{P_{t}}+\lambda\beta^{2}_{l}\mathbf{\hat{H}}^{H}_{1}\mathbf{\hat{H}}_{1}+\lambda\beta^{2}_{l} {\it Tr}\{\mbox{\boldmath$\Sigma$}_{1}\}\mbox{\boldmath$\Psi$}_{1} \Bigg)^{-1}\\&\quad \times\mathbf{\hat{H}}^{H}_{1}\mathbf{T}^{H}_{l}\mathbf{\hat{H}}^{H}_{2},\label{eq:mmseprecoder3}
\end{split}
\end{equation}
\begin{equation}
\beta_{l}=\frac{{\it Tr}\{\Re[\mathbf{P}^{H}_{l}\mathbf{\hat{H}}^{H}_{1}\mathbf{T}^{H}_{l}\mathbf{\hat{H}}^{H}_{2}]\}}{\Omega_1+\lambda \Omega_2}\label{eq:beta1}
\end{equation}
where
\begin{equation}
\begin{split}
\mathbf{M}&=\beta^{2}_{l}\Big(E\big[ \mathbf{\hat{H}}^{H}_{1}\mathbf{T}^{H}_{l}\mathbf{\hat{H}}^{H}_{2}\mathbf{\hat{H}}_{2}\mathbf{T}_{l}\mathbf{\hat{H}}_{1}\big]\\&\quad+{\it Tr}\{\mathbf{\hat{H}}_{2}\mathbf{T}_{l}
\mbox{\boldmath$\Sigma$}_{1}\mathbf{T}^{H}_{l}\mathbf{\hat{H}}^{H}_{2}\} \mbox{\boldmath$\Psi$}_{1}\\&\quad+{\it Tr}\{\mbox{\boldmath$\Sigma$}_{2}\}\mathbf{\hat{H}}^{H}_{1}\mathbf{T}^{H}_{l}
\mbox{\boldmath$\Psi$}_{2}\mathbf{T}_{l}\mathbf{\hat{H}}_{1}\\&\quad+{\it Tr}\{\mbox{\boldmath$\Sigma$}_{2}\} {\it Tr}\{\mathbf{T}^{H}_{l}\mbox{\boldmath$\Psi$}_{2}\mathbf{T}_{l}\mbox{\boldmath$\Sigma$}_{1}\}
\mbox{\boldmath$\Psi$}_{1}\Big),
\end{split}
\end{equation}
 $\Omega_1={\it Tr}\{\mathbf{P}^{H}_{l}\mathbf{\hat{H}}^{H}_{1}\mathbf{D}\mathbf{\hat{H}}_{1}\mathbf{P}_{l}\}+ {\it Tr}\{\mathbf{D}\mbox{\boldmath$\Sigma$}_{1}\} {\it Tr}\{\mathbf{P}^{H}_{l}\mbox{\boldmath$\Psi$}_{1}\mathbf{P}_{l}\}+\sigma^{2}_{1}  {\it Tr}\{\mbox{\boldmath$\Psi$}_{2}\}\mbox{\boldmath$\Sigma$}_{2}+\sigma^{2}_{1} {\it Tr}\{\mathbf{\hat{H}}_{2}\mathbf{\hat{H}}^{H}_{2}\}$, $\Omega_2={\it Tr}\{\mathbf{\hat{H}}_{1}\mathbf{P}_{l}\mathbf{P}^{H}_{l}\mathbf{\hat{H}}^{H}_{1}\}+  {\it Tr}\{\mbox{\boldmath$\Sigma$}_{1}\} {\it Tr}\{\mathbf{P}^{H}_{l}\mbox{\boldmath$\Psi$}_{1}\mathbf{P}_{l}\}+\sigma^{2}_{1}N_{r}$ and $\mathbf{D}=\mathbf{T}^{H}_l\big(\mathbf{\hat{H}}^{H}_{2}\mathbf{\hat{H}}_{2}+{\it Tr}\{\mbox{\boldmath$\Sigma$}_{2}\}\mbox{\boldmath$\Psi$}_{2}\big)\mathbf{T}_l$. The Lagrange multiplier $\lambda$ in (\ref{eq:mmseprecoder3}) and (\ref{eq:beta1}) is given by
\begin{equation}
\lambda=\Bigg[\frac{\Re[\mathbf{P}^{H}_{l}\mathbf{\hat{H}}^{H}_{1}\mathbf{T}^{H}_{l}\mathbf{\hat{H}}^{H}_{2}]\sqrt{\frac{\Omega_2}{P_{r}}} -\Omega_1}{\Omega_2}\Bigg]^{+}.\label{eq:lambda1}
\end{equation}
 The detailed derivation is shown in Appendix \ref{Section8:appendix1}.

 The
solutions for the robust BS precoding matrix and
the RS power scaling factor corresponding to the $l$-th latent precoding matrix pair can be obtained by
implementing (\ref{eq:mmseprecoder3}), (\ref{eq:beta1})  and (\ref{eq:lambda1})
  iteratively
with an initial value of $\mathbf{P}_{l}$. The iterative
optimization algorithm is summarized in Table \ref{tab:table1}.

\fnbelowfloat
 \begin{table*}[t]
\centering
 \caption{\normalsize   The iterative optimization algorithm for the latent precoding matrix pair} {
\begin{tabular}{ll}
\hline
 $1$ & $\mathbf{for}$ the $l$-th latent precoding matrix pair.\\
 $2$ & ~ Initialization: $\mathbf{P}_{l}$.\\
 $3$ & ~ Compute  the Lagrange multiplier $\lambda$ based on (\ref{eq:lambda1}). \\
 $4$ & ~ Compute  the RS power scaling factor $\beta_{l}$ by using (\ref{eq:beta1}).\\
 $5$ & ~ Compute the robust BS precoding matrix $\mathbf{P}_{l}$ based on (\ref{eq:mmseprecoder3}) and $\mathbf{P}_{l}\leftarrow\sqrt{\frac{P_{t}}{ {\it Tr}\{\mathbf{P}_{l}\mathbf{P}^{H}_{l}\}}}\mathbf{P}_{l}$.   \\
 $6$ & ~ Repeat step $3$, step $4$ and step $5$ until $||\mathbf{P}^{i}_{l}-\mathbf{P}^{i-1}_{l}||^2\leq \epsilon $ and \\
   &  ~ $|\beta^{i}_{l}-\beta^{i-1}_{l}|^2\leq \epsilon$, where $\epsilon$ is a predefined threshold value  (e.g. $\epsilon$=0.0001).  \\
   &  ~ $\mathbf{P}^{i}_{l}$ and $\beta^{i}_{l}$ denote $\mathbf{P}_{l}$ and $\beta_{l}$ in the $i$-th iteration, respectively, \\
   &  ~ while  $\mathbf{P}^{i-1}_{l}$ and $\beta^{i-1}_{l}$ denote $\mathbf{P}_{l}$ and $\beta_{l}$  in the ($i-1$)-th iteration, respectively. \\
 $7$ & ~  Obtain the $l$-th latent  precoding matrix pair $\{\mathbf{P}_l, \mathbf{W}_l \}$, where $\mathbf{W}_l=\beta_l\mathbf{T}_l$.\\
 $8$ &  Repeat step   $2$-$7$ until all $2^B$ latent precoding matrix pairs are obtained. \\
\hline
\label{tab:table1}
\end{tabular}
}
\end{table*}
Note that
 after we have obtained all the $2^B$ latent precoding matrix pairs,
 the optimum latent precoding matrix pair  should be chosen according to a  selection mechanism to provide the best performance. In the following, we will focus on the description of the proposed  selection mechanism.

\subsection{  Selection Mechanism}
  Having tried various optimization rules,  the squared Euclidean distance seems to be the best candidate for a simple and yet effective selection mechanism.
Ideally, the optimum latent precoding matrix pair  can be chosen to minimize the accumulated squared Euclidean distance between the true transmit
symbol and the received soft information in one transmission data block.  Note that the proposed algorithm implemented at the BS cannot obtain the exact received signal at the MS, but it has  full information about the transmitted symbols. To overcome this limitation, we propose to use the noiseless information to estimate the received signal. The simulation results in Section \ref{Section6:simulations} verify the effectiveness of the approximation.
Let $\mathbf{s}_{j}$ denote a $KM\times 1$ vector corresponding to the $j$-th transmission data block, which is given by $\mathbf{s}_{j}=[\mathbf{b}^{T}_{(j-1)M+1},\ldots,\mathbf{b}^{T}_{jM}]^{T}$, where $j\in \{1,2,\ldots \}$ and
$M$ is the block length. The $K\times 1$ vector $\mathbf{b}_{(j-1)M+m}=[b_{1},\ldots, b_{K}]^{T}$ denotes the $m$-th transmit vector of the $j$-th block, $
 m \in \{1,\ldots,M\}$. Let $\mathbf{u}^{(l)}_{j}$ denote the $KM\times 1$ pre-estimated vector, which is given by $\mathbf{u}^{(l)}_{j}=[\mathbf{\hat{y}}^{T(l)}_{(j-1)M+1},\ldots,\mathbf{\hat{y}}^{T(l)}_{jM}]^{T}$, where $\mathbf{\hat{y}}^{(l)}_{(j-1)M+m}$ denotes the $K\times 1$ noiseless BS
pre-estimated received vector based on the $l$-th  unitary
matrix for the $m$-th transmit vector of the $j$-th block,  and it is given by
\begin{equation}
\mathbf{\hat{y}}^{(l)}_{(j-1)M+m}=
\beta_{l}\mathbf{\hat{H}}_{2}\mathbf{T}_{l}\mathbf{\hat{H}}_{1}\mathbf{P}_{l}\mathbf{b}_{(j-1)M+m}.\label{eq:preestimate}
\end{equation}
 The optimum  latent precoding matrix pair is chosen based on minimizing the summation of the  squared Euclidean distance values in one transmission data block.
Hence, we have the following selection rule:
\begin{equation}
l_{opt}=\arg \min_{1\leq l\leq
2^B}\bigg \{||\mathbf{s}_{j}-\mathbf{u}^{(l)}_{j}||^2 \bigg\}. \label{eq:selectionsymbol}
\end{equation}
It is worth mentioning that the selection operation takes place once per  block.

\subsection{Codebook Design}

In the following, we introduce a  design method for the codebook of unitary matrices\footnote{According to the discussion of the Grassmannian subspace packing  and  Lloyd algorithms in \cite{love2}-\cite{narula},
the codebook $\mathcal{T}=\{\mathbf{T}_{1},\mathbf{T}_{2},\ldots, \mathbf{T}_{2^{B}}\}$ should be designed such that $\delta=\min_{1\leq l<m\leq 2^{B}}d(\mathbf{T}_{l},\mathbf{T}_{m})$ is as large as possible,
where $d(\mathbf{T}_{l}, \mathbf{T}_{m})=\sqrt{N_{r}-||\mathbf{T}^{H}_{l}\mathbf{T}_{m} ||^{2}_{F}}$.
Note that our proposed algorithm is based on employing unitary matrices as the entries of the codebook.
In this respect, due to $||\mathbf{T}^{H}_{l}\mathbf{T}_{m} ||^{2}_{F}=N_{r}$ for every two given unitary matrices, we have
 $d(\mathbf{T}_{l}, \mathbf{T}_{m})=0$.
Therefore, the Grassmannian or Lloyd algorithm becomes the general  method
that groups $2^{B}$ randomly generated unitary
 matrices to create the  codebook
 and may not be suitable for the proposed SR-based precoding scheme.
}, referred to as   most frequently selected candidates (MSC),
%
 the basic principle of which is to build a
codebook which contains the  unitary matrices for the most
likely selected elements.
To build the codebook, we need to perform an extensive set of
experiments and compute the frequency of the indices of the selected
unitary matrices. Finally, we create the codebook based on the statistics of the indices
and choose the $2^{B}$ candidates  which are  most frequently selected
as  entries of the codebook. The algorithm is summarized
in Table \ref{tab:table3},
\fnbelowfloat
 \begin{table*}[t]
\centering
 \caption{\normalsize  MSC codebook design algorithm} {
\begin{tabular}{ll}
\hline
$1$ & Initialize the vectors $\mathbf{d}$ and $\mathbf{d}_{\textit{idx}}$,
  generate null vectors  for them.
   $\mathbf{d}\leftarrow \mathbf{0}$, $\mathbf{d}_{\textit{idx}}\leftarrow \mathbf{0}$.
   \\& Decide the number of experiments $N_{e}$ and the size of the codebook $2^{B}$.\\
$2$ & Choose an appropriate value for $\alpha$. \\
$3$ & Generate $\alpha$  unitary matrices randomly as the  candidates,
store them in the set $\{\mathbf{F}_{1}, \mathbf{F}_{2}, \ldots, \mathbf{F}_{\alpha}\}$
 \\& and   assign the list of
the  unitary matrices to the vector $\mathbf{d}_{0}$.\\
$4$ & $\mathbf{for}$ $n_{e}=1$ to $N_{e}$ $\mathbf{do}$\\
$5$ & ~ Generate the testing matrices $\mathbf{\hat{H}}_{1}$ and $\mathbf{\hat{H}}_{2}$.\\
$6$ & ~~$\mathbf{for}$ $l=1$ to $\alpha$ $\mathbf{do}$\\
$7$ & ~~~~ Compute the precoding matrix $\mathbf{P}_{l}$ and $\beta_{l}$ based on
$\mathbf{F}_{l}$, $\mathbf{\hat{H}}_{1}$ and $\mathbf{\hat{H}}_{2}$.
\\& ~~~~ Compute the squared Euclidean distance for all the possibilities  \\& ~~~~ of the transmit symbol vector, and assign it to the $l$-th element  of the vector $\mathbf{d}$.\\ & ~~~~  $\mathbf{d}(l)\leftarrow\sum^{\Phi}_{j=1}||\mathbf{b}_{j}-\mathbf{\hat{y}}^{(l)}_{j}||^{2}$.\\
$8$ & ~~$\mathbf{end}$\\
$9$ & ~ Select the  entry corresponding to
 the minimum  squared Euclidean distance \\& ~ from the vector $\mathbf{d}_{0}$
in the $n_{e}$-th experiment,
assign it to the $n_{e}$-th element of the vector $\mathbf{d}_{\textit{idx}}$.
\\& ~ $\mathbf{d}_{\textit{idx}}(n_{e})\leftarrow
\mathbf{MINIndex}(\mathbf{d})$.
\\$10$ & $\mathbf{end}$ \\
$11$ & Based on the vector $\mathbf{d}_{\textit{idx}}$, a histogram
$\mathbf{HIST}(\mathbf{d}_{\textit{idx}})$ is generated.
\\&
   The codebook $\mathcal{T}_{\textit{MSC}}$ is created by selecting the most
   frequently selected $2^{B}$ candidates according to $\mathbf{HIST}(\mathbf{d}_{\textit{idx}})$. \\&
    $\mathcal{T}_{\textit{MSC}}\leftarrow\mathbf{SELECT}(\mathbf{HIST}(\mathbf{d}_{\textit{idx}}))$.\\
\hline
\label{tab:table3}
\end{tabular}
}
\end{table*}
where $\mathbf{d}$ denotes the vector of squared  Euclidean distances for
$\alpha$ possible  unitary matrices. We generate  the
 $\alpha$  unitary matrices randomly,  where $\alpha$ should be a large integer but practical for the experiment, $2^{B}<\alpha$. The quantity $N_{e}$
denotes the total number of experiments, $\mathbf{d}_{\textit{idx}}$
is defined for the storage of the selected candidates for every
experiment. The vector $\mathbf{d}_{0}$ contains the list of  all  $\alpha$ unitary matrices.
The vectors $\{\mathbf{b}_{1},\ldots, \mathbf{b}_{\Phi}\}$ denote all the possibilities of the $K\times 1$ transmit vector. For the case of quadrature phase shift keying (QPSK) modulation, we have $\Phi=4^{K}$ possibilities.
The vector $\mathbf{\hat{y}}^{(l)}_{j}$ denotes the noiseless BS pre-estimated received vector with respect to the $l$-th  unitary matrix and the transmit vector $\mathbf{b}_{j}$, and it is given by $\mathbf{\hat{y}}^{(l)}_{j}=
\beta_{l}\mathbf{\hat{H}}_{2}\mathbf{F}_{l}\mathbf{\hat{H}}_{1}\mathbf{P}_{l}\mathbf{b}_{j}$, $j\in \{1,\ldots, \Phi\}$.
 We highlight that in each run,  after we have computed the squared Euclidean distance
for  all the unitary matrices, the one  which yields the minimum squared Euclidean distance is
stored in $\mathbf{d}_{\textit{idx}}$ at step $9$. Finally, the MSC
codebook $\mathcal{T}_{\textit{MSC}}$ is created by selecting the
most frequently selected $2^{B}$  unitary matrices according to the histogram
of $\mathbf{d}_{\textit{idx}}$. Note that the MSC codebook design method is implemented offline.

\section{Analysis of The Proposed Algorithm}
\label{Section5:analysis2}
In this section, we carry out an analysis of the
proposed algorithm in terms of downlink transmission efficiency, the error probability performance  and  computational complexity.

\subsection{Downlink Transmission Efficiency}

From the aforementioned discussion about the proposed scheme, we know that
for every block prior to payload transmission, there is a preamble
transmission from the BS to the RS which contains  the index of the selected  unitary matrix  and  the RS power scaling factor.
%
We insert the limited feedforward bits at the beginning of the corresponding data block.
The block of the multiantenna scheme comprises $M$ symbol periods each one consisting of $K$ spatial streams, and
the feedforward rate of the optimum index is one per data
block. We use $B$ bits to  represent $2^B$  unitary
matrices and $C$ bits to represent the RS power scaling factor, and assume
that $Q$-ary modulation is  used for the proposed SR-based precoding
scheme,  thus
a number of $B+C$ signalling bits has to be sent for
every $KM\log_{2}(Q)$ transmitted bits in the block.
Note that we rely on a TDD system, transmit CSI can be obtained by exploiting channel reciprocity.
Therefore, the  downlink
transmission efficiency is given by
\begin{equation}
\eta=\frac{KM\log_{2}(Q)}{KM\log_{2}(Q)+B+C}.
\end{equation}
Let us focus on the QPSK modulation and employ a data
block of $M=10$ symbols. For a configuration with
$N_{t}=N_{r}=K=6$, by using $B=C=6$ feedforward bits  we achieve a  downlink
transmission efficiency of $91\%$.
\subsection{Discussion of Error Probability Performance}

In this part, an error probability  performance analysis for our
proposed algorithm is carried out. We divide the problem into two
circumstances based on the side information fed forward from the BS
to the RS, and discuss the performance based on the  total
probability theorem.

It is easy to show that the average error probability over all the destination MSs  can be derived as
\begin{equation}
\mathcal{\bar{P}}_{e}=\frac{1}{K}\sum^{K}_{k=1}\mathcal{P}_{e_{k}},
\end{equation}
where $\mathcal{P}_{e_{k}}$ denotes the  probability of making an error in the symbol
detection for the $k$-th MS. We will rely here on presenting a simple
approach to estimate these probabilities. By using the total
probability theorem, we can write
\begin{equation}
\mathcal{P}_{e_{k}}=\mathcal{P}\{e_{k}|\mathbf{E}^{(k)}_{1}\}\mathcal{P}\{\mathbf{E}^{(k)}_{1}\}+\mathcal{P}\{e_{k}|\mathbf{E}^{(k)}_{2}\}\mathcal{P}\{\mathbf{E}^{(k)}_{2}\},
\end{equation}
where
the  events $\mathbf{E}^{(k)}_{j}$, $j=1,2$, are associated with the  perfect feedforward transmission of side information and the  imperfect feedforward transmission of side information, respectively. They are two mutually exclusive  events, with $\mathcal{P}\Big\{ \cup^{2}_{j=1}\mathbf{E}^{(k)}_{j}\Big\}$=1.

For the event of perfect  side information, we assume that the residual multiuser interference can be approximated as a Gaussian random variable.
%
 In the case with QPSK modulation,
the error probability $\mathcal{P}\{e_{k}|\mathbf{E}^{(k)}_{1}\}$ on the  event $\mathbf{E}^{(k)}_{1}$ can be expressed by
\begin{equation}
\mathcal{P}\{e_{k}|\mathbf{E}^{(k)}_{1}\}=\mathcal{Q}\Big(\sqrt{{\gamma}^{(l_{opt})}_{k}}\Big)
\end{equation}
 where ${\gamma}^{(l_{opt})}_{k}$ denotes the  $k$-th MS's signal-to-interference-plus-noise ratio (SINR) of the optimum latent precoding matrix pair caused by the  unitary matrix $\mathbf{T}_{l_{opt}}$. The structure of ${\gamma}^{(l_{opt})}_{k}$ is given as (\ref{eq:sinranalysis}),
 \fnbelowfloat
\begin{figure*}
\begin{equation}
{\gamma}^{(l_{opt})}_{k}=\frac{\beta^{2}_{l_{opt}}\mathbf{\bar{H}}_{l_{opt}}(k,:)\mathbf{P}_{l_{opt}}(:,k)\mathbf{P}^{H}_{l_{opt}}(:,k)\mathbf{\bar{H}}^{H}_{l_{opt}}(k,:)}{\sum^{K}_{k^{'}\neq k}\big(\beta^{2}_{l_{opt}}\mathbf{\bar{H}}_{l_{opt}}(k,:)\mathbf{P}_{l_{opt}}(:,k)\mathbf{P}^{H}_{l_{opt}}(:,k)\mathbf{\bar{H}}^{H}_{l_{opt}}(k,:)\big)+\beta^{2}_{l_{opt}}\sigma^{2}_{1}\mathbf{H}_{2}(k,:)\mathbf{H}^{H}_{2}(k,:)+\sigma^{2}_{2}},\label{eq:sinranalysis}
\end{equation}
\end{figure*}
where $k \in \{1,\ldots,K\}$.
The function $\mathcal{Q}(.)$ is defined as
the Gaussian error function $\mathcal{Q}(x)=(1/2)\rm erfc (x/\sqrt{2})$.

The  probability $\mathcal{P}\{\mathbf{E}^{(k)}_{1}\}$ relies on the feedforward transmission scheme of side information.   For the case where 
binary PSK modulation is
used in a frequency-nonselective, slow Rayleigh fading channel,
the error probability for each side information bit is given by $\mathcal{P}_{b}=\frac{1}{2}\Big(1-\sqrt{\frac{\Gamma}{1+\Gamma}} \Big)
$   \cite{proakis},
where $\Gamma=\frac{E_{b}}{N_{0}}E[\varphi^{2}]$,
 $\varphi$ represents the Rayleigh-distributed amplitude of the  channel coefficient, $E_{b}$ denotes the energy per bit and $N_{0}$ is the noise power spectral density.   In the event that we transmit $B$ side information bits,
the probability $\mathcal{P}\{\mathbf{E}^{(k)}_{1}\}$ is expressed as
\begin{equation}
\mathcal{P}\{\mathbf{E}^{(k)}_{1}\}=(1-\mathcal{P}_{b})^{B}.
\end{equation}

   In the case of imperfect  side information, the error probability expression of $\mathcal{P}\{e_{k}|\mathbf{E}^{(k)}_{2}\}$ on the  event $\mathbf{E}^{(k)}_{2}$ cannot be derived  due to  misadjustment in the  latent precoding matrix pair selection at the RS.
  However,  in the case that the detection of side information is significantly  affected  by  errors, the selected index of the  latent precoding matrix pair at the BS is not in accordance with the one at the RS. The decision on the preprocessing data becomes random and the error probability  $\mathcal{P}\{e_{k}|\mathbf{E}^{(k)}_{2}\}$ is $0.5$. Following the above example, the error probability $\mathcal{P}_{e_{k}}$ for the $k$-th MS is given by $\mathcal{P}_{e_{k}}=0.5(1-(1-\mathcal{P}_{b})^B)+\mathcal{Q}\Big(\sqrt{{\gamma}^{(l_{opt})}_{k}}\Big)(1-\mathcal{P}_{b})^{B}$,
where the probability of imperfect side information transmission $\mathcal{P}\{\mathbf{E}^{(k)}_{2}\}$ is  $1-(1-\mathcal{P}_{b})^{B}$.
  We remark that an accurate error probability expression of $\mathcal{P}_{e_{k}}$ cannot be obtained as a result of the specific nature of the proposed scheme. It remains an open problem.
In section \ref{Section6:simulations}, we will illustrate the error probability performance in the presence of side information errors.

\subsection{ Computational Complexity}

We measure the complexity in terms of the number of  floating point (FLOP).
From \cite{ggolub}, we know that  a complex addition and multiplication has $2$ and $6$ FLOPs, respectively.
We note that the complexity of the matrix inversion is cubic in the number of BS or RS antennas \cite{Horn}.
In Table \ref{tab:complexity}, we show the complexity
of the conventional precoding algorithm, the proposed latent precoding matrix pair design algorithm and  the
 selection mechanism of the proposed scheme.
The overall complexity of the proposed
algorithm includes the complexity of the selection mechanism and the design complexity of  each latent precoding matrix pair multiplied by the codebook size  $2^B$. The complexity of the proposed algorithm increases with the codebook size.
%
In the simulation section, we will show that for a limited increase in complexity the performance of the proposed SR-based robust precoding design algorithm outperforms the performance of the conventional precoding algorithms significantly.
In practice, the codebook size should be chosen to achieve a suitable trade-off between performance requirements and implementation complexity, based on  a given
channel environment.

\fnbelowfloat
\begin{table*}[t]
\centering%
\caption{\normalsize   Computational complexity} {
\begin{tabular}{cc}
\hline   $\mathbf{Algorithms}$  & $\mathbf{Complexity}$ \\
\hline
 \emph {Robust Relay Precoding }\cite{Binzhang}  & $\mathcal{O}[N^3_{r}+KN^2_{r}+N^2_{r}+N_{r}]$ \\
\hline
\hline
 \emph{ SR-based Precoding (each iteration)}: &\\
  precoding matrix (\ref{eq:mmseprecoder3})  &  $\mathcal{O}[N^3_{t}+(N_{r}+K)N^2_{t}+(N_{t}+K)N^2_{r}+K^2+N_{r}N_{t}K+KN^2_{t}+N_{t}K]$  \\
   $\beta_{l}$  &  $\mathcal{O}[N^3_{r}+KN^2_{r}+N_{r}K^2+N_{t}K^2]$  \\
\hline
 \emph {  Selection Mechanism} & $\mathcal{O}[\big((K+N_{r})N_{t}+N^2_{r}+KN_{r}+K\big)M+KM]$ \\
  \hline
\end{tabular}
}\label{tab:complexity}
\end{table*}

\section{ Simulation results}
\label{Section6:simulations}
In this section, we conduct simulations to
evaluate the  proposed SR-based robust precoding scheme and compare it  with
  existing precoding algorithms for
multiuser MIMO relaying systems \cite{Binzhang}-\cite{zijianwang}.
In the simulations, we assume that both the first phase MIMO channel and the second
phase MIMO broadcast channel are quasi-static flat fading channels
 with a Rayleigh distribution. $10000$ channel realizations are
employed for each simulation.
The configuration
of the system is $N_{t}=N_{r}=K=6$.
By using the exponential model \cite{chengwenxing1}, \cite{minghuadingtsp2009} and \cite{xizhangtsp2008}, the channel estimation error covariance matrices can be expressed as
\begin{equation}
{\bf{\Psi }}_{1}  = {\bf{\Psi }}_{2}  = \begin{bmatrix}
1 & \theta & \theta ^2 & \theta ^3 & \theta^4 & \theta^5 \\
\theta & 1 & \theta  & \theta ^2 & \theta^3 & \theta^4 \\
\theta ^2 & \theta  & 1 & \theta & \theta^2 & \theta^3 \\
\theta ^3 & \theta ^2 & \theta & 1 & \theta & \theta^2 \\
\theta^4 & \theta^3 & \theta^2 & \theta & 1 & \theta \\
\theta^5 & \theta^4 & \theta^3 & \theta^2 & \theta & 1 \\
\end{bmatrix},\quad
\end{equation}
\begin{equation}
  {\bf{\Sigma }}_{1}  =\sigma _e^2 \begin{bmatrix}
1 & \rho & \rho ^2 & \rho ^3 & \rho^4 & \rho^5 \\
\rho & 1 & \rho  & \rho ^2 & \rho^3 & \rho^4 \\
\rho ^2 & \rho  & 1 & \rho & \rho^2 & \rho^3 \\
\rho ^3 & \rho ^2 & \rho & 1 & \rho & \rho^2 \\
\rho^4 & \rho^3 & \rho^2 & \rho & 1 & \rho \\
\rho^5 & \rho^4 & \rho^3 & \rho^2 & \rho & 1\\
\end{bmatrix},\quad
{\bf{\Sigma }}_{2}=\sigma _e^2 \mathbf{I},
\end{equation}
where $\theta$ and $\rho$ denote the correlation coefficients, and $\sigma^{2}_{e}$ is the estimation error variance.
Since the destination MSs are  far apart spaced and uncorrelated, the
correlation coefficient in  the covariance matrix of the second
phase channel seen from the receiver is zero. The estimated
channels, $\hat{\mathbf{H}}_{j}$, $j=1, 2$, are therefore
generated based on the following   distribution:
\begin{equation}
 \hat{\mathbf{H}}_{j} \sim\mathcal{CN} \left( {{\bf{0}} ,\frac{{\left( {1 - \sigma _e^2 } \right)}}{{\sigma _e^2 }}{\bf{\Sigma }}_{j}  \otimes {\bf{\Psi }}_{j}^T } \right)
\end{equation}
 such that the channel realizations  $\mathbf{H}_{j}=\mathbf{\hat{H}}_{j}+\Delta\mathbf{H}_{j}$  have unit variance.
We set $P_{t}=P_{r}=K$,
%
and define the input SNR$=P_{t}/\bar{\sigma}^{2}$, where
$\bar{\sigma}=\sigma_{1}=\sigma_{2}$.
 In the simulations, the BS employs $C=6$ bits to quantize the computed RS power scaling parameter.
With regard to the scalar information, we use a nonuniform scalar quantizer \cite{jayant}.
This information is fed forward to the RS, together with
the $B$ bits corresponding to the index of the selected latent precoding matrix pair.
%
The iterative optimization algorithm for each latent precoding matrix pair uses the identity matrix as the initial value of the precoding matrix.
 QPSK modulation is used as the modulation scheme.
 Among the analyzed techniques in this paper, we consider the following:
\begin{enumerate}[$\bullet$]
\item SR precoding: the proposed SR-based robust precoding algorithm.

\begin{enumerate}[$1)$]

\item $B$-bit: the limited feedforward schemes employ $B$ bits corresponding to the index of the selected  latent precoding matrix pair, namely $2^{B}$ is the codebook size.

\item MSC: the proposed SR-based  precoding scheme with the codebook generated by the MSC method.

\item Random: the proposed SR-based  precoding scheme with $2^B$
randomly generated unitary matrices in the codebook.

\end{enumerate}

\item Robust Identity:
The BS precoding matrix is designed based on the robust MMSE technique with the conventional relay scheme, which amplifies the energy of the received data at the RS and
forwards the signal directly \cite{Laneman2}.  That is to say, the RS precoding matrix is an identity matrix.

\item Robust Relay MMSE: the MMSE-based robust MIMO RS precoding algorithm proposed in \cite{Binzhang}\footnote{Although it is developed based on the feedback quantized channel errors, we have extended the algorithm straightforwardly to the case with channel estimation errors for the comparison.}.

\item SVD-ZF: the SVD-based joint BS and RS ZF precoding algorithm proposed in \cite{weixu1}, which is only based on the estimated CSI.

\item SVD-RZF: the SVD-based robust joint BS and RS ZF precoding algorithm proposed in \cite{ zijianwang}.

\end{enumerate}



\begin{figure}[!hhh]
\centering
\scalebox{0.63}{\includegraphics{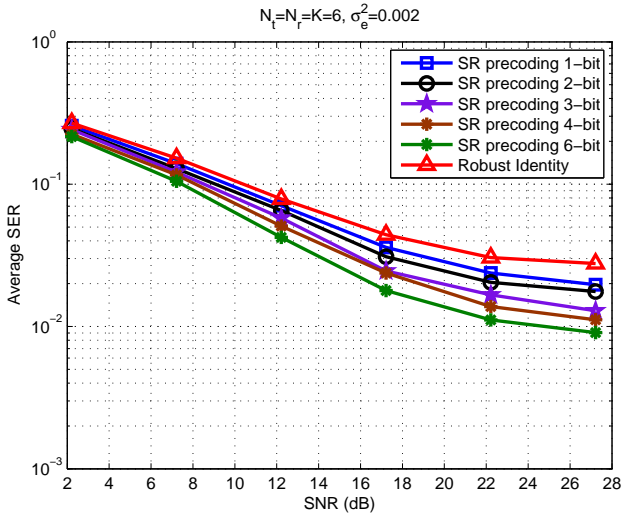}}
\caption{ Average SER performance versus SNR for the proposed SR-based robust precoding scheme. $\sigma^{2}_{e}=0.002$, $\theta=\rho=0$.
} \label{fig:fig3}
\end{figure}

Fig. \ref{fig:fig3} shows the average SER performance versus input SNR for the proposed SR-based precoding scheme, i.e. $1$-, $2$-, $3$-, $4$-, and $6$-bit for the index of the selected  latent precoding matrix pair, respectively. We apply the MSC method for the  unitary matrix codebook design.  We set $N_{e}=10000$ and $\alpha=1000$.
The channel estimation error variance is given by $\sigma^{2}_{e}=0.002$, and the correlation coefficients are given by $\theta=\rho=0$. The channel varies per transmission data block, each block contains $M=10$ symbols.
From the results, we can see that the best performance is achieved with the proposed scheme with $B=6$  bits, and the average SER decreases as the number of feedforward bits increases. In the simulation,  we assume that  perfect side information is fed forward.

\begin{figure}[!hhh]
\centering
\scalebox{0.63}{\includegraphics{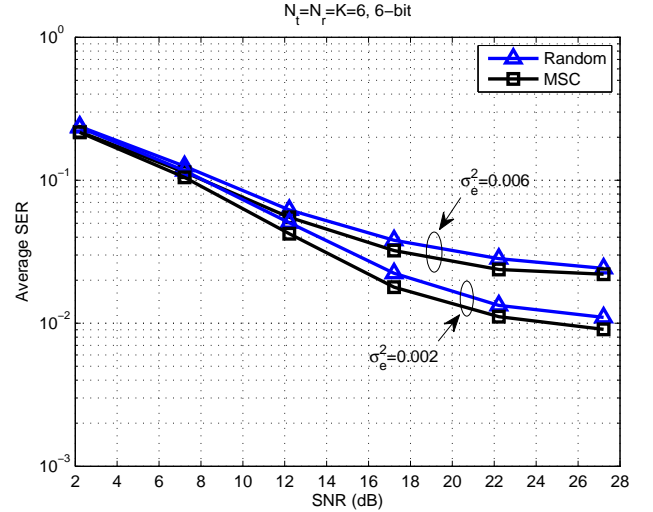}}
\caption{ Average SER performance versus SNR for the proposed SR-based robust precoding scheme.
$B=6$, $N_{e}=10000$, $\alpha=1000$. $\theta=\rho=0$.
} \label{fig:fig2}
\end{figure}

We compare the codebooks of the
 unitary matrices which are created  by  two methods, namely
the randomly generated method and the proposed MSC
method. In particular, we show average SER performance curves
versus input SNR for different values of  estimation error variance. Note that the codebooks are
designed offline. For the MSC algorithm we set the number
of simulation $N_{e} = 10000$ and the number of candidates
$\alpha= 1000$.   The channel coefficients
 are generated independently.  The results which are illustrated in Fig. \ref{fig:fig2} show that the performance of the proposed precoding scheme with different codebooks of  unitary matrices, where we use $B=6$ bits.
We can see that the MSC method outperforms the random method.
Compared to the   random method, the proposed MSC codebook design
method  can have  a gain of $2$ dB.

\begin{figure}[!hhh]
\centering
\scalebox{0.63}{\includegraphics{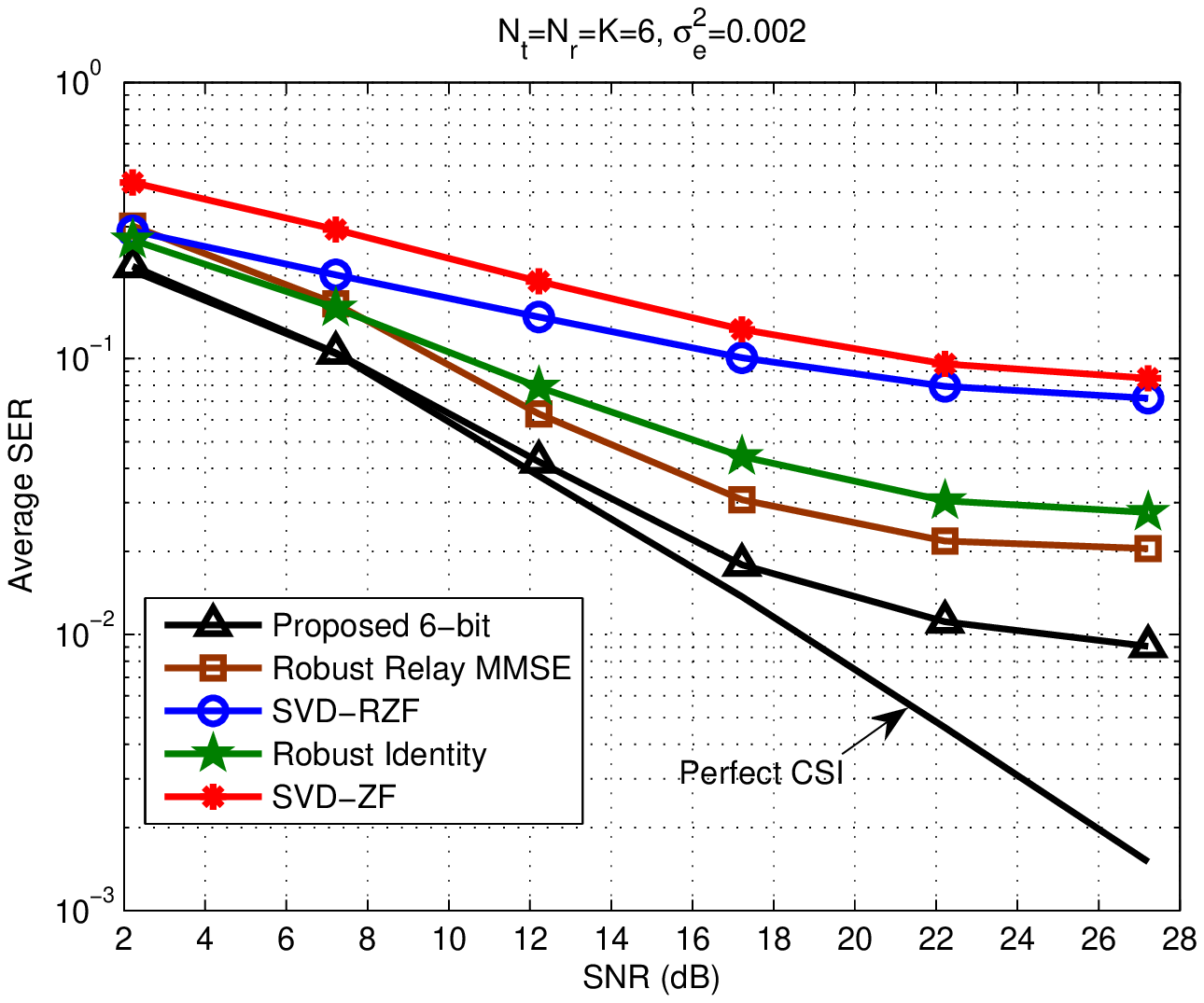}}
\caption{ Average SER performance versus SNR for the proposed SR-based robust precoding scheme and the existing relay precoding schemes. $\sigma^{2}_{e}=0.002$, $\theta=\rho=0$, $B=6$.
} \label{fig:fig4}
\end{figure}

\begin{figure}[!hhh]
\centering
\scalebox{0.63}{\includegraphics{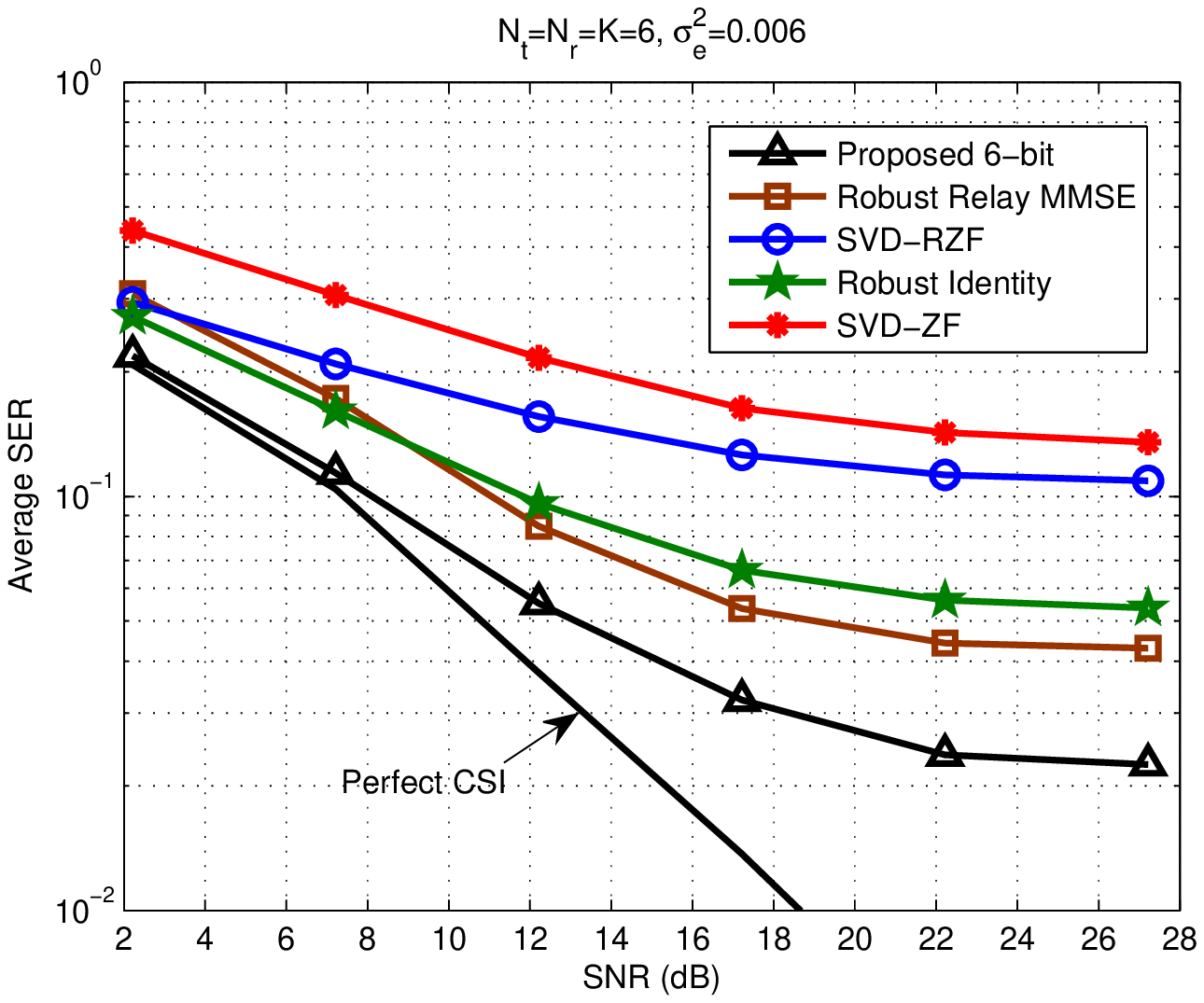}}
\caption{ Average SER performance versus SNR for the proposed SR-based robust precoding scheme and the existing relay precoding schemes. $\sigma^{2}_{e}=0.006$, $\theta=\rho=0$, $B=6$.
} \label{fig:fig5}
\end{figure}

Fig. \ref{fig:fig4} and Fig. \ref{fig:fig5} compare the average SER
versus  the SNR  of the proposed SR-based precoding scheme
with some existing relay precoding algorithms. The MSC method is
used for the codebook design. The same system configuration and
channel model are employed here.
In Fig. \ref{fig:fig4}, the channel error variance is given by $\sigma^{2}_{e}=0.002$, and the  correlation coefficients are given by $\theta=\rho=0$.
The performance of the proposed SR-based robust MMSE precoding scheme is much better than the others. In particular, the proposed SR-based robust precoding scheme with $B=6$ bits  can save over  $3$ dB in transmit power in comparison with the robust relay MMSE precoding algorithm,  at an average SER level of $2 \times 10^{-2}$.
The SER performance of the SR-based precoding scheme with $B=6$ bits under perfect CSI is given as a reference.
In Fig. \ref{fig:fig5}, the channel error variance is given by
$\sigma^{2}_{e}=0.006$, and the  correlation coefficients are given
by $\theta=\rho=0$. We can see that the best performance is achieved
with the proposed SR-based robust precoding scheme with $B=6$ bits,
followed by the robust relay MMSE precoding algorithm, the robust identity
technique, the SVD-RZF precoding algorithm, and the  SVD-ZF
precoding algorithm. Specifically,  at an average
SER level of $5\times 10^{-2}$ the proposed SR-based robust
precoding scheme can save $5$ dB in comparison with the robust relay
MMSE precoding algorithm. The results show the ability of the proposed SR-based
precoding algorithm to handle channel uncertainties and multiuser
interference.


\begin{figure}[!hhh]
\centering
\scalebox{0.63}{\includegraphics{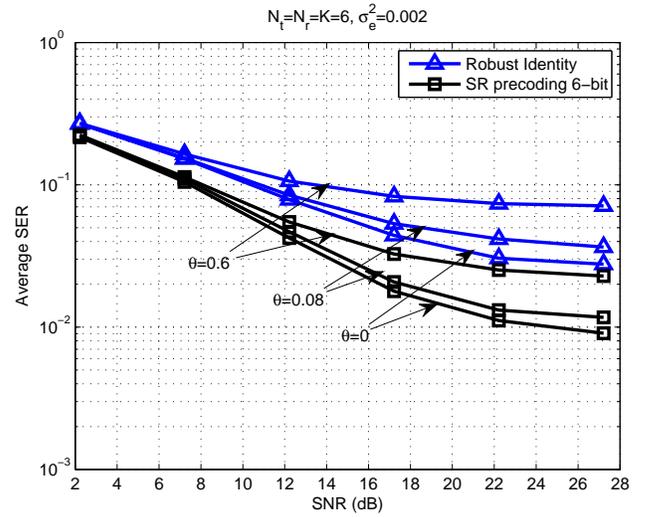}}
\caption{ Average SER performance versus SNR for the proposed SR-based robust precoding and robust identity schemes. $B=6$, $\sigma^{2}_{e}=0.002$, $\rho=0$.
} \label{fig:fig7}
\end{figure}

\begin{figure}[!hhh]
\centering
\scalebox{0.63}{\includegraphics{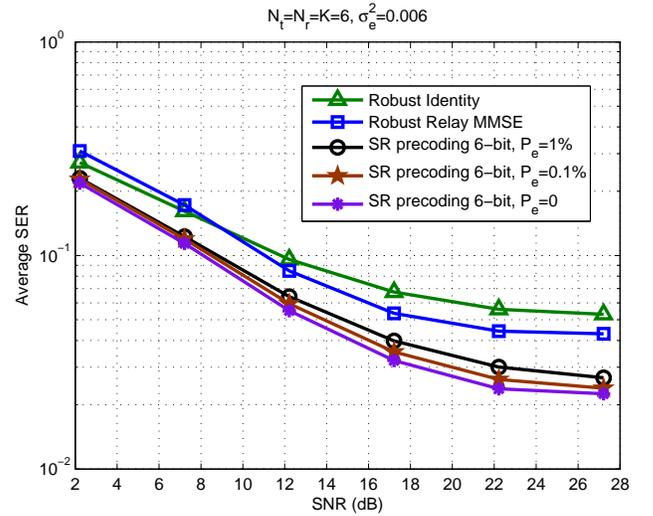}}
\caption{ Average SER performance versus SNR for the proposed SR-based robust precoding and conventional schemes. $B=6$, $\sigma^{2}_{e}=0.006$, $\theta=\rho=0$.
} \label{fig:fig6}
\end{figure}
Fig. \ref{fig:fig7} shows the SER  performance comparison for
the proposed SR-based precoding scheme and the existing robust
identity technique with different values of $\theta$. In this
simulation, we let $\rho=0$, $\sigma^{2}_{e}=0.002$  and $\theta$ was
varied. From Fig. \ref{fig:fig7}, it can be seen that smaller
correlation coefficients lead to a better performance.
 When the value of $\theta$ decreases, the performance of both algorithms improve. The performance of the proposed SR precoding algorithm with $B=6$ bits is always superior to the performance of the conventional robust identity algorithm. In particular, the proposed SR precoding algorithm can save up to  $10$ dB in transmit power in comparison with the robust identity algorithm,  at an SER level of $4\times 10^{-2}$  in the case with $\theta=0.08$.  This demonstrates the ability of the proposed  algorithm to properly handle CSI uncertainty as well as  channel correlation.



 In the next simulation, we focus on  examining
the performance of the proposed algorithm in the presence of feedforward side information errors.
The last results, shown in Fig. \ref{fig:fig6}
illustrate the averaged SER performance with different levels of  side information  errors
for the proposed SR-based precoding scheme.
We use a structure based on a frame format where the indices are
converted to $0$s and $1$s. This frame of $1$s and $0$s with the
feedforward information is transmitted over a binary symmetric channel associated
with a probability of error $P_{e}$.
The burst error scenario in the limited feedforward channel can be easily transferred to the case of the binary symmetric channel by employing a conventional bit interleaver.
In particular,
we use $B=6$  bits for the  index of the selected  unitary matrix and $C=6$ bits to represent  the computed RS power scaling factor.
We let $\rho=\theta=0$ and $\sigma^{2}_{e}=0.006$.
The MSC method is  used for the codebook design. The same system configuration and channel model
are employed here.
As we increase the
feedforward side information errors, the performance of the proposed limited feedforward
scheme decreases, since the  unitary matrices at  the BS and RS are not equal to each other due to  feedforward errors.
 Associated with a side information  error level of $P_{e}=0.1\%$,
  the performance has  $1$dB degradation,  compared with the perfect side information case at a BER level of $3\times 10^{-2}$.
%
In order to guarantee that the errors are controlled,   channel coding
techniques should be  used for the signalling feedforward channels with large
errors.




\section{Conclusion}
\label{Section7:conclusion}

In this paper, we have proposed a
robust MMSE BS precoding strategy based on SR processing for multiuser MIMO relaying systems.
We have also developed a
selection mechanism, which was used for symbol detection.
A  method based on the  most selected candidates for the  unitary matrix codebook design has been proposed.
 We have discussed the error probability, the computational complexity and the transmission efficiency of the proposed scheme and algorithms.
%
 The results have shown that the proposed
SR-based scheme significantly outperforms the existing relay precoding algorithms in the presence of imperfect CSI.
 Our future work will extend our proposed algorithms to take into account  systems with other precoding schemes.

\appendices
\section{Derivation for (\ref{eq:mmseprecoder3}), (\ref{eq:beta1}) and (\ref{eq:lambda1})}
\label{Section8:appendix1}
  By focusing on the RS transmit power constraint,  we obtain the following Lagrangian objective function\footnote{In this work, we simply scale the computed BS precoding expression to meet the BS transmit power constraint \cite{ADDabbagh}.}:
\begin{equation}
\begin{split}
J(\mathbf{P}_{l},\beta_{l},\lambda)&=\zeta+\lambda\Big(\beta^{2}_{l} E\Big[ {\it Tr}\{\mathbf{H}_{1}\mathbf{P}_{l}\mathbf{P}^{H}_{l}\mathbf{H}^{H}_{1}+\sigma^{2}_{1}\mathbf{I}\}\Big]-P_{r}\Big)\label{eq:lag155}
\end{split}
\end{equation}
 where  $\lambda$ denotes the Lagrange multiplier for the RS transmit power constraint.  Based on
 the KKT conditions, we have:
 \begin{equation}
 \beta^{2}_{l} E\Big[ {\it Tr}\{\mathbf{H}_{1}\mathbf{P}_{l}\mathbf{P}^{H}_{l}\mathbf{H}^{H}_{1}+\sigma^{2}_{1}\mathbf{I}\}\Big]- P_{r} \leq 0,
 \end{equation}
  \begin{equation}
 \lambda\Big(\beta^{2}_{l}E\Big[ {\it Tr}\{\mathbf{H}_{1}\mathbf{P}_{l}\mathbf{P}^{H}_{l}\mathbf{H}^{H}_{1}+\sigma^{2}_{1}\mathbf{I}\}\Big]- P_{r}\Big) = 0, \quad \lambda\geq 0 \label{eq:lag355}
 \end{equation}
 \begin{equation}
 \bigtriangledown  J(\mathbf{P}_{l},\beta_{l},\lambda)_{\mathbf{P}^{*}_{l}}=\mathbf{0}, \quad \bigtriangledown  J(\mathbf{P}_{l},\beta_{l},\lambda
  )_{\beta_{l}}=0.
 \end{equation}
 The RS transmit power is given by
 \begin{equation}
 \begin{split}
 \beta^{2}_{l} E\Big[ {\it Tr}\{\mathbf{H}_{1}\mathbf{P}_{l}\mathbf{P}^{H}_{l}\mathbf{H}^{H}_{1}+\sigma^{2}_{1}\mathbf{I}\}\Big]&=\beta^{2}_{l}{\it Tr}\{\mathbf{\hat{H}}_{1}\mathbf{P}_{l}\mathbf{P}^{H}_{l}\mathbf{\hat{H}}^{H}_{1}\\&\quad+E\big[\Delta\mathbf{H}_{1}\mathbf{P}_{l}\mathbf{P}^{H}_{l}\Delta\mathbf{H}^{H}_{1}\big]\\&\quad+\sigma^{2}_{1}\mathbf{I}\}.
\end{split}
\end{equation}
By taking the gradient terms of (\ref{eq:lag155}) with respect to $\mathbf{P}^{*}_{l}$ and  equating them to zero, we  can obtain (\ref{eq:mmseprecoder3}).

By taking the gradient terms of (\ref{eq:lag155}) with respect to $\beta_{l}$ and  equating them to zero, we obtain
\begin{equation}
\begin{split}
\bigtriangledown
J(\mathbf{P}_{l},\beta_{l},\lambda)_{\beta_{l}}&=2\beta_{l}E\Big[ {\it Tr}\{\mathbf{\bar{H}}_{l}\mathbf{P}_{l}\mathbf{P}^{H}_{l}\mathbf{\bar{H}}^{H}_{l}\}\Big]\\&\quad+2\sigma^{2}_{1} \beta_{l} E\Big[ {\it Tr}\{\Delta\mathbf{H}_{2}\Delta\mathbf{H}^{H}_{2}\}\Big]\\&\quad+2\sigma^{2}_{1}\beta_{l} {\it Tr}\{\mathbf{\hat{H}}_{2}\mathbf{\hat{H}}^{H}_{2}\}\\&\quad+2\lambda\beta_{l} E\Big[ {\it Tr}\{\mathbf{H}_{1}\mathbf{P}_{l}\mathbf{P}^{H}_{l}\mathbf{H}^{H}_{1}+\sigma^{2}_{1}\mathbf{I}\}\Big]\\&\quad- {\it Tr}\{\mathbf{P}^{H}_{l}\mathbf{\bar{H}}^{H}_{l}+\mathbf{\bar{H}}_{l}\mathbf{P}_{l}\}=0.\label{eq:lambda0055}
\end{split}
\end{equation}
By solving (\ref{eq:lambda0055}) we have (\ref{eq:beta1}).

By substituting (\ref{eq:beta1}) into (\ref{eq:lag355}) and solving the equation, we have (\ref{eq:lambda1}).
In order to meet the transmit
power constraint $ {\it Tr}\{\mathbf{P}_{l}\mathbf{P}^{H}_{l}\}=P_{t}$, the proposed BS precoding matrix for the $l$-th
latent precoding matrix pair is given by $\mathbf{P}_{l}\leftarrow\sqrt{\frac{P_{t}}{ {\it Tr}\{\mathbf{P}_{l}\mathbf{P}^{H}_{l}\}}}\mathbf{P}_{l} $,
where the arrow denotes an overwrite operation.


\begin{thebibliography}{1}


%





%




%
%






\bibitem{yunlongwcnc2013}
Y. Cai, R. C. de Lamare, L.-L. Yang and M. Zhao, ``Robust MMSE precoding strategy for multiuser MIMO relay systems with switched relaying and side information," in  \textit{Proc. IEEE Wireless Communications and Networking Conference (WCNC)}, Apr. 2013, Shanghai, China.

\bibitem{WGUAN}
W. Guan and H. Luo, ``Joint MMSE transceiver design in
non-regenerative MIMO relay systems,"  \textit{IEEE Commun.
Lett.,} vol. 12, no. 7, pp. 517-519, Jul. 2008.




\bibitem{yuerong1}
Y. Rong and F. Gao, ``Optimal beamforming for non-regenerative MIMO relays with direct link,"  \textit{IEEE Commun. Lett.,} vol. 13, no. 12, pp. 926-928, Dec. 2009.


\bibitem{chengwenxing1}
C. Xing, S. Ma and Y.-C. Wu, ``Robust joint design of linear relay precoder and destination equalizer for dual-hop amplify-and-forward MIMO relay systems," \textit{ IEEE Trans. Signal Process.} vol. 58, no. 4, pp. 2273-2283, Apr. 2010.

\bibitem{Binzhang2}
B. Zhang, X. Wang, K. Niu and Z. He, ``Joint linear transceiver design for non-regenerative MIMO relay systems,"  \textit{Elect. Lett.,} vol. 45, no. 24, pp. 1254-1256, Nov. 2009.

\bibitem{fanshuotseng}
F. Tseng and W. Wu, ``Linear MMSE transceiver design in amplify-and-forward MIMO relay systems,"  \textit{IEEE Trans. Veh. Technol.,} vol. 59, no. 2, pp. 754-765, Feb. 2010.

\bibitem{fanshuotseng2}
F. Tseng, W. Wu, and J. Wu, ``Joint source/relay precoder design in nonregenerative cooperative systems using an MMSE criterion,"  \textit{IEEE Trans. Wireless Commun.,} vol. 8, no. 10, pp. 4928-4933, Oct. 2009.









\bibitem{Rzhang}
R. Zhang, C. C. Chai, and Y. C. Liang, ``Joint beamforming and power
control for multiantenna relay broadcast channel with QoS
constraints,"  \textit{IEEE Trans. Signal Process.,} vol. 57, no.
2, pp. 726-737, Feb. 2009.

\bibitem{Njindal}
%
C. Chae, T. Tang, R. W. Heath, Jr., and S. Cho, ``MIMO relaying with linear processing for multiuser transmission in fixed relay networks,"  \textit{IEEE Trans. Signal Process.,} vol. 56, no. 2, pp. 727-738, Feb. 2008.

\bibitem{Binzhang}
B. Zhang, Z. He, K. Niu and L. Zhang, ``Robust linear beamforming
for MIMO relay broadcast channel with limited feedback,"
\textit{IEEE Signal Process. Lett.,} vol. 17, no. 2, pp. 209-212, Feb. 2010.

%



\bibitem{weixu1}
W. Xu, X. Dong, and W.-S. Lu, ``Joint optimization for source and relay precoding under multiuser MIMO downlink channels," in \textit{Proc. IEEE International Conference on Communications (ICC)}, May 2010.

\bibitem{weixu2}
W. Xu, X. Dong, and W.-S. Lu, ``MIMO relaying broadcast channels with linear precoding and quantized channel state information feedback,"  \textit{IEEE Trans. Signal Process.,} vol. 58, no. 10, pp. 5233-5245, Oct. 2010.

\bibitem{zijianwang}
Z. Wang, W. Chen, and J. Li, ``Efficient beamforming for MIMO relaying broadcast channel with imperfect channel estimation,"  \textit{IEEE Trans. Veh. Technol.,} vol. 61, no. 1, pp. 419-426, Jan. 2012.

\bibitem{keke2}
K. Zu, R. C. de Lamare, ``Low-Complexity Lattice Reduction-Aided
Regularized Block Diagonalization for MU-MIMO Systems'', IEEE.
Communications Letters, Vol. 16, No. 6, June 2012, pp. 925-928.

\bibitem{keke1} K. Zu, R. C. de Lamare and M. Haart,
``Generalized design of low-complexity block diagonalization type
precoding algorithms for multiuser MIMO systems", IEEE Trans.
Communications, 2013.

\bibitem{zuthp}
K. Zu, R. C. de Lamare and M. Haardt, ``Multi-Branch
Tomlinson-Harashima Precoding Design for MU-MIMO Systems: Theory and
Algorithms," IEEE Transactions on Communications, vol.62, no.3,
pp.939,951, March 2014.

\bibitem{rmbthp}
L. Zhang, Y. Cai, R. C. de Lamare and M. Zhao, "Robust Multibranch
Tomlinson–Harashima Precoding Design in Amplify-and-Forward MIMO
Relay Systems," IEEE Transactions on Communications, vol.62, no.10,
pp.3476,3490, Oct. 2014.

\bibitem{Dfeng1}
D. Feng, C. Jiang, G. Lim, L. J. Cimini, Jr., G. Feng, and G. Y. Li, ``A Survey of Energy-Efficient Wireless Communications," \textit{IEEE Commun. Surveys  Tutorials,} vol. 15, no. 1, pp. 167-178, First Quarter 2013.

\bibitem{gyli}
G. Y. Li, Z. Xu, C. Xiong, C. Yang, S. Zhang, Y. Chen, and S. Xu, ``Energy-Efficient Wireless Communications: Tutorial, Survey, and Open Issues," \textit{IEEE Wireless Commun.,} pp. 28-35, Dec. 2011.

\bibitem{mBanerjee}
N. Banerjee, M. D. Corner, D. Towsley, and B. N. Levine, ``Relays, Base Stations, and Meshes: Enhancing Mobile Networks with Infrastructure," in \textit{Proc. MOBICOM,} San Francisco, USA., Sep. 2008.

\bibitem{vook}
F. W. Vook, X. Zhuang, K. L. Baum, T. A. Thomas, and M. C. Cudak, ``Signaling methodologies to support closed-loop transmit processing in TDD-OFDMA," IEEE C802.16e-04/103r2, Jul. 2004.

%
%
\bibitem{ycai}
Y. Cai, R. C. de Lamare and R. Fa, ``Switched
interleaving techniques with limited feedback for interference
mitigation in DS-CDMA systems," \textit{IEEE Trans. Commun.},
vol. 59, no. 7, pp. 1946-1956, Jul. 2011.

\bibitem{mcsi}
Y. Cai, R. C. de Lamare and D. Le Ruyet, ``Transmit Processing
Techniques Based on Switched Interleaving and Limited Feedback for
Interference Mitigation in Multiantenna MC-CDMA Systems," IEEE
Transactions on Vehicular Technology, vol.60, no.4, pp.1559,1570,
May 2011.

\bibitem{CBchae}
C. Chae, D. Mazzarese, T. Inoue, and R. W. Heath, Jr., ``Coordinated beamforming for the multiuser MIMO broadcast channel with limited feedforward,"  \textit{IEEE Trans. Signal Process.,} vol. 56, no. 12, pp. 6044-6056, Dec. 2008.


%


%
%
%
%
%

 \bibitem{swpeters}
 S. W. Peters, A. Y. Panah, K. T. Truong, and R. W, Heath Jr., ``Relay architectures for 3GPP LTE-Advanced,"  \textit{EURASIP J. on Wireless Commun. and Networking,}
vol. 2009, Article ID 618787.



\bibitem{Laneman2}
J. N. Laneman and G. W. Wornell, ``Cooperative diversity in wireless network: Efficient protocols and outage behaviour,"  \textit{IEEE
Trans. Inf. Theory,} vol. 50, no. 12, pp. 3062-3080, Dec. 2004.

%
%



%

\bibitem{musavian}
L. Musavian, M. Nakhai, M. Dohler, and A. Aghvami, ``Effect of channel uncertainty on the mutual information of MIMO fading channels," \textit{IEEE Trans. Veh. Technol.,} vol. 56,
no. 5, pp. 2798-2806, Sep. 2007.

\bibitem{ding2}
M. Ding and S. Blostein, ``MIMO minimum total MSE transceiver design
with imperfect CSI at both ends," \textit{IEEE Trans. Signal
Process.,} vol. 57, no. 3, pp. 1141-1150, Mar. 2009.

\bibitem{delamare_spa}
R. C. de Lamare, R. Sampaio-Neto, \emph{Minimum Mean-Squared Error
Iterative Successive Parallel Arbitrated Decision Feedback Detectors
for DS-CDMA Systems}, \hskip 1em plus 0.5em minus 0.4em\relax IEEE
Trans. on Commun., vol. 56, p.p. 778-789, May 2008.

\bibitem{LL11_TWC}
P. Li, R. C. de Lamare and R. Fa, ``Multiple Feedback Successive
Interference Cancellation Detection for Multiuser MIMO Systems,"
\textit{IEEE Transactions on Wireless Communications}, vol. 10, no.
8, pp. 2434 - 2439, August 2011.

\bibitem{mdfpic}
P. Li and R. C. de Lamare, "Adaptive Decision-Feedback Detection
With Constellation Constraints for MIMO Systems", \emph{IEEE
Transactions on Vehicular Technology}, vol. 61, no. 2, 853-859,
2012.

\bibitem{mbdf}
R. C. de Lamare, "Adaptive and Iterative Multi-Branch MMSE Decision
Feedback Detection Algorithms for Multi-Antenna Systems", \emph{IEEE
Transactions on Wireless Communications}, vol. 14, no. 10, October
2013.


\bibitem{Tkong}
T. Kong and Y. Hua, ``Optimal design of source and relay pilots for
MIMO relay channel estimation," \textit{IEEE Trans. Signal
Process.}, vol. 59, no. 9, pp. 4438-4446, Sep. 2011.

\bibitem{P.Cla}
P. Clarke ad R. C. de Lamare, ``Transmit Diversity and Relay
Selection Algorithms for Multi-relay Cooperative MIMO Systems'',
IEEE Transactions on Vehicular Technology, vol. 61 , no. 3, March
2012, pp. 1084 - 1098.

\bibitem{Wang}
T. Wang, R. C. de Lamare, and P. D. Mitchell, "Low-Complexity
Set-Membership Channel Estimation for Cooperative Wireless Sensor
Networks," \emph{IEEE Trans. Veh. Technol.}, vol. 60, no. 6, May,
2011.

\bibitem{SSUN}
 S. Sun and Y. Jing,``Channel training design in amplify-and-forward MIMO relay networks," \textit{IEEE Trans. Wireless Commun.}, vol. 10, no. 10, pp. 3380-3391, Oct. 2011.

 \bibitem{jpais_iet}
R. C. de Lamare, ``Joint iterative power allocation and linear
interference suppression algorithms for cooperative DS-CDMA
networks", IET Communications, vol. 6, no. 13 , 2012, pp. 1930-1942.

\bibitem{TARMO}
T. Peng, R. C. de Lamare and A. Schmeink, ``Adaptive Distributed
Space-Time Coding Based on Adjustable Code Matrices for Cooperative
MIMO Relaying Systems'', \emph{IEEE Transactions on Communications},
vol. 61, no. 7, July 2013.

\bibitem{Tseng2013tvt}
F. Tseng, W. Huang and W. Wu, ``Robust Far-End Channel Estimation in
Three-Node Amplify-and-Forward MIMO Relay Systems," \textit{IEEE
Trans. Veh. Technol.,} vol. 62, no. 8, pp. 3752-3766, Oct. 2013.

\bibitem{jidf}
R. C. de Lamare and R. Sampaio-Neto, ``Adaptive Reduced-Rank
Processing Based on Joint and Iterative Interpolation, Decimation,
and Filtering," \textit{IEEE Transactions on Signal Processing},
vol. 57,  no. 7,  July 2009, pp. 2503 - 2514.

\bibitem{delamaretvt10}
R. C. de Lamare and R. Sampaio-Neto, ``Reduced-Rank Space-Time
Adaptive Interference Suppression With Joint Iterative Least Squares
Algorithms for Spread-Spectrum Systems," \textit{IEEE Transactions
on Vehicular Technology}, vol.59, no.3, March 2010, pp.1217-1228.

\bibitem{jiomimo}
R. C. de Lamare and R. Sampaio-Neto, ``Adaptive reduced-rank
equalization algorithms based on alternating optimization design
techniques for MIMO systems", \textit{IEEE Transactions on Vehicular
Technology}, vol. 60, no. 6, 2482-2494, 2011.



%
%

\bibitem{james}
A. T. James, ``Distributions of matrix variates and latent roots derived from normal samples," \textit{Ann. Math. Stat.,} vol. 35, pp. 475-501, 1964.

\bibitem{cxing2}
C. Xing, S. Ma, Z. Fei, Y.-C. Wu and H. V. Poor, ``A General Robust Linear Transceiver Design for Multi-Hop Amplify-and-Forward MIMO Relaying Systems," \textit{IEEE Trans. Signal Process.,} vol. 61, no. 5, pp. 1196-1209, Mar. 2013.


%

%

\bibitem{Palomar2010}
 J. Wang and D. P. Palomar, ``Robust MMSE Precoding in MIMO Channels With Pre-Fixed Receivers," \textit{IEEE Trans. Signal Process.,} vol. 58, no. 11, pp. 5802-5818, Nov. 2010.

\bibitem{ADDabbagh}
A. D. Dabbagh and D. J. Love, ``Multiple antenna MMSE based
downlink precoding with quantized feedback or channel mismatch,"  \textit{IEEE Trans. Commun.,} vol. 56, no. 11, pp.
1859-1868, Nov. 2008.

\bibitem{Sboyd}
S. Boyd and L. Vandenberghe, \textit{Convex Optimization.} Cambridge
University press, 2004.

\bibitem{love2}
D. J. Love and R. W. Heath, Jr., ``Limited feedback unitary precoding for orthogonal space-time block codes," \textit{IEEE Trans. Signal Process.,} vol. 53, no. 1, pp. 64-73, Jan. 2005.


\bibitem{mukkavilli}
K. K. Mukkavilli, A. Sabharwal, E. Erkip, and B. Aazhang, ``On beamforming with finite rate feedback in multiple-antenna systems," \textit{IEEE Trans. Inf. Theory,} vol. 49, no. 10, pp. 2562-2579, Oct. 2003.

\bibitem{narula}
A. Narula, M. J. Lopez, M. D. Trott, and G. W. Wornell, ``Efficient use of side information in multiple-antenna data transmission over fading channels," \textit{IEEE J. Sel. Areas Commun.,} vol. 16, no. 8, pp. 1423-1436, Oct. 1998.


%

%
%
%




 \bibitem{proakis}
 J. G. Proakis, \textit{Digital Communications,} Electrical Engineering, McGraw-Hill, New York, NY, USA, 5th edition, 2007. pp. 846-847.

\bibitem{ggolub}
G. Golub and C. Van Loan, \textit{Matrix Computations.} Johns Hopkins Univ. Press, 1996.

\bibitem{Horn}
 R. A. Horn and C. R. Johnson, \textit{Matrix Analysis}. Cambridge, U.K.:
Cambridge Univ. Press, 1985.

\bibitem{minghuadingtsp2009}
 M. Ding and S. D. Blostein, ``MIMO Minimum Total MSE Transceiver Design With Imperfect CSI at Both Ends," \textit{IEEE Trans. Signal Process.,} vol. 57, no. 3, pp. 1141-1150, Mar. 2009.


\bibitem{xizhangtsp2008}
 X. Zhang, D. P. Palomar and B. Ottersten, ``Statistically Robust Design of Linear MIMO Transceivers," \textit{IEEE Trans. Signal Process.,} vol. 56, no. 8, pp. 3678-3689, Aug. 2008.

\bibitem{jayant}
N. S, Jayant and P. Noll, \textit{Digital Coding of Waveforms: Principles and Applications to Speech and Video.} Englewood Cliffs, NJ: Prentice-Hall, 1984.



%

%








%







\end{thebibliography}
\end{document}